\documentclass[aps,amsmath,amssymb,twocolumn,showpacs,footinbib]{revtex4}
\usepackage{graphicx}
\usepackage{amssymb}

\begin{document}
\title{Meissner screening mass 
in two-flavor quark matter at nonzero temperature }
\author{O. Kiriyama}
\email{kiriyama@th.physik.uni-frankfurt.de}
\affiliation{Institut f\"ur Theoretische Physik, 
J.W.\ Goethe-Universit\"at, D-60438 Frankfurt am Main, Germany}

\begin{abstract}
We calculate the Meissner screening mass of gluons 4--7 in two-flavor 
quark matter at nonzero temperature. To this end, 
we study the effective potential of the 2SC/g2SC phases 
including a vector condensate $\langle gA_z^6 \rangle$ and 
calculate the Meissner mass from the potential curvature 
with respect to $\langle gA_z^6 \rangle$. 
We find that the Meissner mass becomes real 
at the critical temperature which is about 
the half of the chemical potential mismatch. 
The phase diagram of the neutral two-flavor color superconductor 
is presented in the plane of temperature and coupling strength. 
We indicate the unstable region for gluons 4--7 on the phase diagram.
\end{abstract}
\date{\today}
\maketitle

During the last decade, significant advances have been made 
in our understandings of color superconductivity. 
The studies of QCD-motivated effective theories 
has revealed a rich phase structure of neutral quark matter \cite{CSC}. 
However, the question which phase is realized in nature 
is still inconclusive because of a chromomagnetic instability 
\cite{Huang2004,Casalbuoni2004,Alford2005,Fukushima2005}. 
(For recent discussions on this issue, 
see also Refs. \cite{Giannakis2004,Huang2005,Hong2005,Gorbar2005,Gorbar2005b,Iida2006,
Fukush2006,GHMS2006,Hashimoto2006,KRS2006,Giannakis2006}.) 

In the two-flavor case (and at zero temperature), 
while the 8th gluon has a tachyonic Meissner mass 
in the region $\Delta/\delta\mu < 1$ 
($\Delta$ is a diquark gap and $\delta\mu$ denotes the chemical 
potential mismatch between up and down quarks), 
the more severe instability related to gluons 4--7 emerges 
in the region $\Delta/\delta\mu < \sqrt{2}$ 
(i.e. not only in the g2SC phase but also in the 2SC phase) \cite{Huang2004}. 

At nonzero temperature, the instability is expected 
to be weakened by thermal effects \cite{Alford2005,Fukushima2005}. 
In this work, we calculate 
the Meissner screening mass of gluons 4--7 in two-flavor quark matter 
at nonzero temperature that has not been addressed so far. 
The results are useful for the phase diagram of QCD 
and compact star phenomenology.

To study the Meissner screening mass of gluons 4--7, 
we use a gauged Nambu--Jona-Lasinio (NJL) model 
with massless up and down quarks:
\begin{eqnarray}
{\cal L}&=&\bar{\psi}(iD\hspace{-7pt}/+\hat{\mu}\gamma^0)\psi
+G_D\left(\bar{\psi}i\gamma_5\varepsilon\epsilon^bC\bar{\psi}^T\right)
\left(\psi^TCi\gamma_5\varepsilon\epsilon^b\psi\right)\nonumber\\
&&-\frac{1}{4}F_{\mu\nu}^{a}F^{a\mu\nu},
\end{eqnarray}
where the quark field $\psi$ carries flavor ($i,j=1,\ldots N_f$ 
with $N_f=2$) and color ($\alpha,\beta=1,\ldots N_c$ with $N_c=3$) 
indices, $C$ is the charge conjugation matrix; 
$(\varepsilon)^{ik}=\varepsilon^{ik}$ and 
$(\epsilon^b)^{\alpha\beta}=\epsilon^{b\alpha\beta}$ 
are the antisymmetric tensors in flavor and color spaces, 
respectively. The covariant derivative and the field-strength tensor 
are defined as
\begin{subequations}
\begin{eqnarray}
D_{\mu} &=& \partial_{\mu}-igA_{\mu}^{a}T^{a},\\
F_{\mu\nu}^{a} &=& \partial_{\mu}A_{\nu}^{a}-\partial_{\nu}A_{\mu}^{a}
+gf^{abc}A_{\mu}^{b}A_{\nu}^{c}.
\end{eqnarray}
\end{subequations}
In order to evaluate loop diagrams we use a three-momentum cutoff 
$\Lambda=653.3$ MeV. This specific choice does not affect 
the qualitative features of the present analysis. 

In NJL-type models, one has to impose the neutrality conditions 
by adjusting the values of 
an electron chemical potential $\mu_e$ 
and a color chemical potential $\mu_8$ \cite{BubSho}. 
We neglect the color chemical potential $\mu_8$ throughout 
because it is suppressed in the 2SC/g2SC phases, $\mu_8\ll\Delta$. 
Then, the elements of the diagonal matrix of 
quark chemical potentials $\hat{\mu}$ 
in $\beta$-equilibrated neutral quark matter are given by 
$\mu_{u}=\bar{\mu}-\delta\mu$ and $\mu_{d}=\bar{\mu}+\delta\mu$ 
with $\bar{\mu}=\mu-\delta\mu/3$ and $\delta\mu=\mu_e/2$. 

In Nambu-Gor'kov space, the inverse full quark propagator 
$S^{-1}(p)$ is written as
\begin{eqnarray}
S^{-1}(p)=\left(
\begin{array}{cc}
(S_0^+)^{-1} & \Phi^- \\
\Phi^+ & (S_0^-)^{-1} 
\end{array}
\right),
\end{eqnarray}
with
\begin{subequations}
\begin{eqnarray}
&&(S_0^+)^{-1}=\gamma^{\mu}p_{\mu}+(\bar{\mu}-\delta\mu\tau^3)\gamma^0
+g\gamma^{\mu}A_{\mu}^{a}T^{a},\\
&&(S_0^-)^{-1}=\gamma^{\mu}p_{\mu}-(\bar{\mu}-\delta\mu\tau^3)\gamma^0
-g\gamma^{\mu}A_{\mu}^{a}T^{aT},
\end{eqnarray}
\end{subequations}
and
\begin{eqnarray}
\Phi^- = -i\varepsilon\epsilon^b\gamma_5\Delta~,\qquad
\Phi^+ = -i\varepsilon\epsilon^b\gamma_5\Delta.
\end{eqnarray}
Here $\tau^3=\mbox{diag}(1,-1)$ is a matrix in flavor space. 
We have chosen the diquark condensate to point 
in the third direction in color space. 
In this work we are interested 
in the Meissner screening mass of gluons 4--7, 
so it is sufficient to study the case of 
the nonvanishing vector condensate 
$B\equiv\langle gA_z^6 \rangle \neq 0$ \cite{Gorbar2005}.

In the one-loop approximation, the effective potential 
of two-flavor quark matter with electrons at finite temperature $T$ is given by
\begin{eqnarray}
V(\Delta,B,\delta\mu,\mu,T)&=&-\frac{1}{12\pi^2}
\left(\mu_e^4+2\pi^2T^2\mu_e^2+\frac{7\pi^4}{15}T^4\right)\nonumber\\
&&+\frac{\Delta^2}{4G_D}
-\frac{T}{2}\sum_{n=-\infty}^{\infty}
\int^{\Lambda}\frac{d^3p}{(2\pi)^3}\nonumber\\
&&\times\ln\mbox{Det}S^{-1}(i\omega_n,\vec{p}),
\label{eqn:omega}
\end{eqnarray}
where $\omega_n=(2n+1)\pi T$ are 
the Matsubara frequencies ($n=0,\pm 1,\pm 2,\dots$).

Let us look at the Meissner screening mass 
derived from the effective potential (\ref{eqn:omega}). 
Although the potential curvature reproduce the correct 
hard-dense-loop (HDL) result 
for the Meissner mass squared at zero temperature \cite{Gorbar2005,KRS2006},
\begin{eqnarray}
&&\frac{\partial^2 V(\Delta,B,\delta\mu,\mu,0)}
{\partial B^2}\bigg{|}_{B=0}\nonumber\\
&&=\frac{\bar{\mu}^2}{6\pi^2}
\Bigg[1-\frac{2\delta\mu^2}{\Delta^2}
+2\frac{\delta\mu\sqrt{\delta\mu^2-\Delta^2}}{\Delta^2}
\theta(\delta\mu-\Delta)\Bigg],\label{eqn:mm}
\end{eqnarray}
this is not the case for the Meissner mass squared 
calculated directly from Eq. (\ref{eqn:omega}). 
[Note that terms of order ${\cal O}(\bar{\mu}^2/\Lambda^2)$ 
and ${\cal O}(\Delta^2/\bar{\mu}^2)$ 
have been neglected in Eq. (\ref{eqn:mm}).] 
The potential curvature suffers 
from ultraviolet divergences $\propto\Lambda^2$ 
and therefore we have to subtract them.

At zero temperature the subtraction term is given by
\begin{eqnarray}
\frac{\partial^2 V(0,B,0,0,0)}
{\partial B^2}\bigg{|}_{B=0}=-\frac{\Lambda^2}{3\pi^2}.\label{eqn:cz}
\end{eqnarray}
At nonzero temperature, since we use a finite cutoff, 
the subtraction term should depend on temperature. Indeed one finds
\begin{eqnarray}
&&\frac{\partial^2 V(0,B,\delta\mu,\mu,T)}
{\partial B^2}\bigg{|}_{B=0}\nonumber\\
&&=\frac{\Lambda^2}{12\pi^2}\sum_{e1,e2,e3=\pm}
e_1 N_F(e_1\Lambda+e_2\bar{\mu}+e_3\delta\mu),\label{eqn:ct}
\end{eqnarray}
where $N_F(x)=1/(e^{x/T}+1)$, 
otherwise the Meissner mass in the normal phase 
assumes an unphysical positive value. 
Note that, in the case of $T \to 0$, 
Eq. (\ref{eqn:ct}) is in exact agreement with Eq. (\ref{eqn:cz}). 
It should be also noted that the temperature dependence 
in Eq. (\ref{eqn:ct}) is made redundant 
when $\Lambda \gg T$. 
Thus, the temperature dependence of Eq. (\ref{eqn:ct}) 
is a cutoff artifact indeed. 
%
% Fig. 1
%
\begin{figure}
\includegraphics[width=0.48\textwidth]{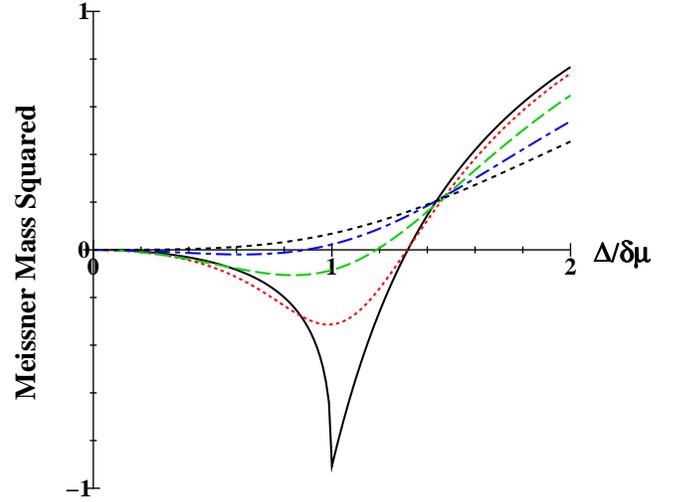}
\caption{The Meissner mass squared 
[divided by $\bar{\mu}^2/(6\pi^2)$] 
as a function of $\Delta/\delta\mu$ for 
$T=0$ MeV (solid), $T=10$ MeV (dotted) $T=20$ MeV (dashed), 
$T=30$ MeV (dot-dashed) $T=40$ MeV (short-dashed). 
We used $\bar{\mu}=500~{\rm MeV}$ and $\delta\mu=80~{\rm MeV}$.}
\label{Figure1}
\end{figure}

Figure \ref{Figure1} shows the Meissner mass squared,
\begin{eqnarray}
m_M^2 &=& \frac{\partial^2}{\partial B^2}
\big[V(\Delta,B,\delta\mu,\mu,T)\nonumber\\
&&\qquad-V(0,B,\delta\mu,\mu,T)\big]\bigg{|}_{B=0},\label{eqn:MM6}
\end{eqnarray}
as a function of $\Delta/\delta\mu$ for several temperatures. 
The results are plotted for $\bar{\mu}=500$ MeV and 
$\delta\mu=80$ MeV. Note that we have solved neither the gap equation 
nor the neutrality condition, thus $\Delta/\delta\mu$ 
in Fig. \ref{Figure1} are temperature independent parameters. 

At $T=0$, we see the manifestation 
of the chromomagnetic instability at all values below 
$\Delta/\delta\mu=\sqrt{2}$. 
(Since our model parameters do not correspond to the HDL limit, 
strictly speaking, the actual critical value of $\Delta/\delta\mu$ 
is somewhat smaller than $\sqrt{2}$.) 
As $T$ is increased, due to thermal smoothing effects, 
the Meissner mass squared tends to approach its value in the normal phase. 
However, its temperature dependence 
at fixed $\Delta/\delta\mu$ shows a non-monotonic behavior. 
This is analogous to the result for a charged condensate 
reported in Ref. \cite{Alford2005}.

The Meissner mass squared at small 
but nonzero values of $\Delta/\delta\mu$ remains 
negative at higher temperatures. 
At $T^{\star} \simeq \delta\mu/2 \simeq 40$ MeV, 
the imaginary Meissner mass 
completely disappears for all values of $\Delta/\delta\mu$ 
\footnote{We have examined the robustness 
of the relation $T^{\star} \simeq \delta\mu/2$ 
varying our model parameters and confirmed that it works well 
as far as $\bar{\mu}$ is not too small.}. 
As $T$ is further increased, 
the Meissner mass squared at nonzero values of $\Delta/\delta\mu$ 
begins to become small and approaches zero. 
(Note that, due to the normalization (\ref{eqn:MM6}), 
$m_M^2$ in the normal phase, $\Delta/\delta\mu=0$, is always zero.)

%
% Fig. 2
%
\begin{figure*}
\includegraphics[width=0.85\textwidth]{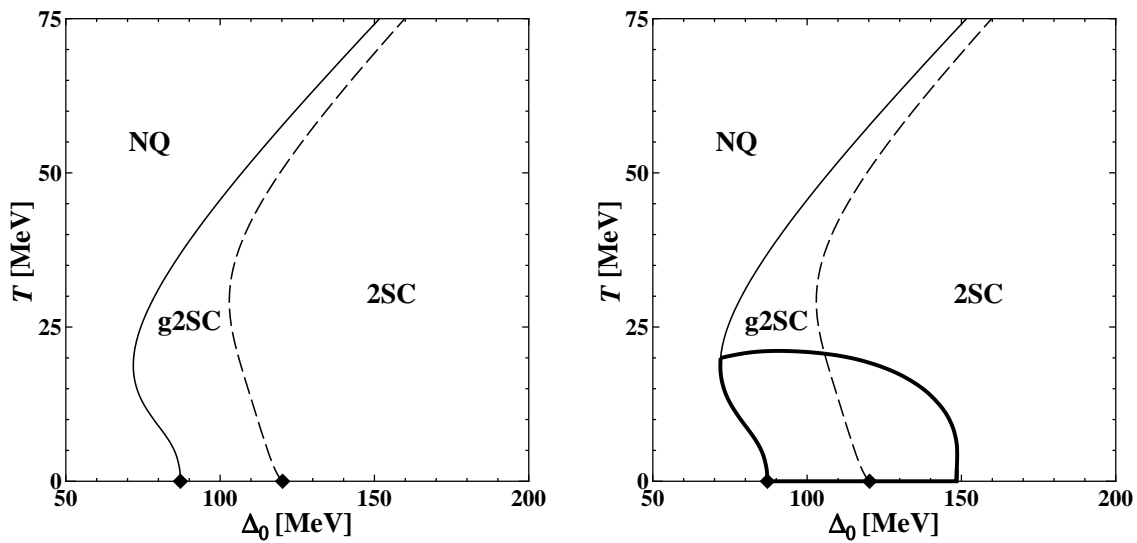}
\caption{Left: The phase diagram of neutral two-flavor color superconductor 
as a function of $T$ and $\Delta_0$. 
The solid (dashed) line denotes the critical line of 
the phase transition between the normal quark phase and the g2SC phase 
(the g2SC phase and the 2SC phase). The results are plotted for $\mu=400$ MeV. 
Right: The same as the left panel, but the unstable region for 
gluons 4--7 is enclosed by the thick solid line.}
\label{Figure2}
\end{figure*}
%
% Fig. 3
%
\begin{figure}
\includegraphics[width=0.48\textwidth]{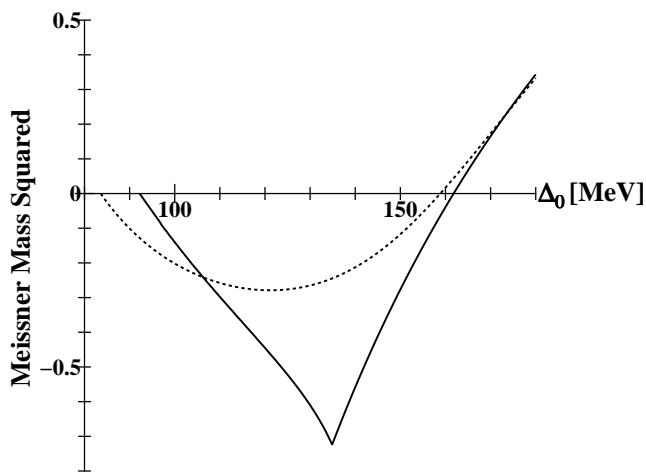}
\caption{The Meissner mass squared [divided by $\bar{\mu}^2/(6\pi^2)$] 
as a function of $\Delta_0$ for 
$T=0$ MeV (solid) and $T=10$ MeV (dotted). 
The results are plotted for $\mu=400$ MeV. At $T=0$ MeV, the curve 
has a kink at the onset of the g2SC phase.}
\label{Figure3}
\end{figure}
In order to investigate the implications of our analysis 
for the phase structures of QCD, 
we solved the gap equation and the neutrality condition:
\begin{subequations}
\label{eqn:ceq}
\begin{eqnarray}
&&\frac{\partial V(\Delta,0,\delta\mu,\mu,T)}{\partial \Delta}=0,\\
&&\frac{\partial V(\Delta,0,\delta\mu,\mu,T)}{\partial \mu_e}=0.
\end{eqnarray}
\end{subequations}
The phase diagram of the neutral 2SC/g2SC matter is displayed 
in Fig. \ref{Figure2} as a function of of $T$ and $\Delta_0$. 
Here, $\Delta_0$ is the value of the 2SC gap at $\delta\mu=0$ 
and at $T=0$. 
The result is plotted for $\mu=400$ MeV, 
which should be relevant to the interior of compact stars. 
The solid (dashed) line denotes the critical line of 
the phase transition between the normal quark phase and the g2SC phase 
(the g2SC phase and the 2SC phase). 
At $T=0$, the g2SC phase exists in the window
\begin{eqnarray}
92~{\rm MeV} < \Delta_0 < 134~{\rm MeV},
\end{eqnarray}
and the 2SC window is given by
\begin{eqnarray}
\Delta_0 > 134~{\rm MeV}.
\end{eqnarray}
These results agree with those obtained by using an approximation 
for the effective potential \cite{Gorbar2005b,KRS2006}. 
As we will see below, 
the phase diagram in the left panel of Fig. \ref{Figure2} 
is qualitatively consistent with those presented 
in the literature \cite{Shovkovy2003,phased}. 

In the weak coupling regime, $77~{\rm MeV}<\Delta_0<92~{\rm MeV}$, 
the chemical potential mismatch is too large 
for these couplings for diquark pairing 
and the system is in the normal quark (NQ) phase at $T=0$. 
At $T>0$, the mismatch of the Fermi surfaces is thermally smeared 
and, then, it opens the possibility of finding the g2SC phase. 
In the intermediate coupling regime, 
$92~{\rm MeV}<\Delta_0<134~{\rm MeV}$, 
we find the g2SC phase even at $T=0$. 
For relatively strong couplings, 
$110~{\rm MeV}<\Delta_0<134~{\rm MeV}$, 
the g2SC phase at low temperature is replaced by the 2SC phase 
at intermediate temperature. 
At higher temperature, the 2SC phase is replaced by the g2SC phase again. 
It is known that this unusual behavior happens 
in the intermediately coupled 
two-flavor color superconductor \cite{Shovkovy2003}. 
For strong coupling, $\Delta_0>134~{\rm MeV}$, 
the gap $\Delta$ increases and the 2SC phase is favored at $T=0$. 
At higher temperatures, however, $\Delta$ is decreased by thermal effects 
and the g2SC phase becomes possible. 

Let us now take into account the chromomagnetic instability. 
Combining Eq. (\ref{eqn:ceq}) with the Meissner mass squared (\ref{eqn:MM6}), 
we calculated the Meissner mass squared 
and mapped out the unstable region for gluons 4--7 
on the $T$-$\Delta_0$ phase diagram. 
The region enclosed by the solid thick line in the right panel 
of Fig. \ref{Figure2} corresponds to the region 
where gluons 4--7 have a negative Meissner mass squared. 
In this region, the gluonic phase is energetically more favored than the 2SC/g2SC phases 
and resolves the instability. 
(In this paper, we refer to the phase where $B \neq 0$ as the gluonic phase. 
Strictly speaking, the reduced symmetry of the 2SC/g2SC phases with $B \neq 0$ 
does not exclude additional gluonic condensates \cite{Gorbar2005}.)

At $T=0$, we find the manifestation of the instability 
in the region $92~{\rm MeV}<\Delta_0<162~{\rm MeV}$ 
(see Fig. \ref{Figure3}) 
\footnote{The upper boundary of 
this unstable region, $\Delta_0=162~{\rm MeV}$, is 
lower than that quoted in Refs. \cite{Gorbar2005b,KRS2006}. 
As mentioned earlier, the fact that 
our parameters do not correspond to the HDL limit yields the difference. 
To find the unstable window we directly calculated 
the Meissner mass squared 
and the resulting critical value of $\Delta/\delta\mu$ is 
much smaller than the HDL value $\sqrt{2}$.}. 
For extremely strong couplings, $\Delta_0 \agt 162$ MeV, 
the 2SC phase is free from the instability 
not only at $T=0$ but also at $T>0$. 
On the other hand, the whole g2SC phase 
and the part of the 2SC phase 
suffer from the instability at low temperatures. 
At $T \simeq 20$ MeV, 
the phase transition from the gluonic phase to the 2SC/g2SC phases 
occurs and the unstable region disappears. 
(The order of the phase transition is most probably second-order \cite{IP}.) 
As is clear from the right panel of Fig. \ref{Figure2}, 
in the g2SC window, the critical temperature for the gluonic $\to$ g2SC 
reaches about half that for the g2SC $\to$ NQ transition. 
As a consequence, 
most of the g2SC phase should be replaced by the gluonic phase. 
In addition, it is very likely that the g2SC phase 
is also unstable for the 8th gluon at $T \ge 0$ \cite{Alford2005}. 
(Our preliminary study shows that, for intermediate couplings, 
the unstable region for the 8th gluon 
survives for temperatures of order of 10 MeV \cite{IP}.) 
Thus, the g2SC phase suffers from the severe instability related 
to gluons 4--7 and 8. 
It should be emphasized, however, that the stable g2SC phase 
is not excluded from the phase diagram. 
Even if we take account of the instability for the 8th gluon, 
we still find the stable g2SC phase 
in high-temperature regions of the phase diagram \cite{IP}, 
though the gapless structure is not so important at nonzero temperature. 

In this paper, we studied the Meissner mass squared 
in the 2SC/g2SC phases at nonzero temperature. 
It was found that the negative Meissner 
mass squared for gluons 4--7 completely 
disappears at the temperature $T^{\star}\simeq\delta\mu/2$. 
We also investigated the $T$-$\Delta_0$ phase diagram 
and mapped out the unstable region for gluons 4--7 on the phase diagram. 
We did not vary the quark chemical potential $\mu$, 
so we could not map out unstable regions explicitly 
on the $T$-$\mu$ phase diagram. 
However, the result is an indication that wide regions 
of the g2SC phase on the phase diagram 
suffer from the instability. 

Resolving the chromomagnetic instability 
is the most pressing task 
in the study of the phase diagram of color super conductor. 
It is , therefore, interesting to reconsider 
the $T$-$\mu$ phase diagram of QCD, 
taking into account a LOFF state 
with realistic crystal structures \cite{ABR,Rajagopal2006}, 
a colored crystalline phase \cite{Fukushima2005}, 
a phase with a vector condensate \cite{Huang2005,Hong2005} 
and gluonic condensates \cite{Gorbar2005}. 
Such a study would have important implications 
for the physics of compact stars

\begin{acknowledgments}
The author would like to thank Igor Shovkovy for fruitful discussions 
and Dirk Rischke for comments on the manuscript. 
This work was supported by the Deutsche Forschungsgemeinschaft (DFG).

{\it Note added.} 
After finishing this work, I learned that an overlapping study 
was recently done by L. He, M. Jin, and P. Zhuang \cite{He}.
\end{acknowledgments}

\end{document}